\documentclass{pos}

\usepackage{bm}
\usepackage{axodraw}
\usepackage{bbm}

\input pix.sty

\def\Scat{\picc{%
 \Laqu(10,-2.5)(50,-2.5)%
 \Lagh(50,-2.8)(50,17.8)%
 \Lqu(10,17.5)(50,17.5)%
 \Lgh(10,-2.8)(10,17.8)%
}}
\def\ScatA{\picc{%
 \Laqu(20,-2.5)(50,-2.5)%
 \Line(10,-2.5)(20,-2.5)%
 \Lagh(50,-2.8)(50,17.8)%
 \Lqu(10,17.5)(40,17.5)%
 \Line(40,17.5)(50,17.5)%
 \Lgh(10,-2.8)(10,17.8)%
 \Lgl(30,-2.5)(30,17.5)%
}}
\def\ScatB{\picc{%
 \Laqu(10,-2.5)(50,-2.5)%
 \Lagh(50,-2.8)(50,17.8)%
 \Lqu(10,17.5)(50,17.5)%
 \Lgh(10,-2.8)(10,17.8)%
 \Agl(30,17.5)(6,180,360)%
}}
\def\ScatC{\picc{%
 \Laqu(10,-2.5)(50,-2.5)%
 \Lagh(50,-2.8)(50,17.8)%
 \Lqu(10,17.5)(50,17.5)%
 \Lgh(10,-2.8)(10,17.8)%
 \Agl(30,-2.5)(6,0,180)%
}}
\title{Heavy Quarkonia beyond Deconfinement and Real Time Lattice Simulations }

\ShortTitle{Heavy Quarkonia beyond Deconfinement and Real Time Lattice Simulations }

\author{Marcus Tassler
         \thanks{The work presented in this article was done in collaboration with Mikko Laine and Owe Philipsen under support of the BMBF project \textit{Hot Nuclear Matter from Heavy Ion Collisions and its Understanding from QCD}. I wish to thank Paul Romatschke as a collaborator in an earlier stage of the project.}
	\\
        University of Muenster\\
        E-mail: \email{marcus.tassler@uni-muenster.de}}

\abstract{Since the initial investigation by Matsui and Satz heavy quark bound states at finite temperature have been subject to numerous studies. The derivation of a finite-temperature potential from first principles was attempted only recently however, by generalising the Schr\"odinger equation which is successfully employed for the description of quarkonia at zero temperature to a thermal setting. In this note the finite-temperature static potential is derived to leading order using resummed perturbation theory. The modification of the heavy quarkonium spectral function by an imaginary part of the potential appearing at finite temperature is discussed. Additionally, the extent of possible corrections due to non-perturbative processes is assessed by employing real-time lattice techniques based on kinetic theory.}

\FullConference{8th Conference Quark Confinement and the Hadron Spectrum\\
		 September 1-6, 2008\\
		 Mainz. Germany}

\begin{document}

\section{Introduction}
\noindent
The purpose of this brief note is to present a recently suggested definition of the finite-temperature static potential from first principles, which can be directly related to the quarkonium spectral function. This so called \textit{real-time static potential} \cite{Laine:2006ns} is calculated using resummed perturbation theory to leading order. The method of real-time lattice simulations is introduced to analyse corrections due to non-perturbative physics present in the classical limit of the theory. For related work on this subject see also \cite{Brambilla:2008cx,Beraudo:2008fx}. 
\section{Real-time static potential}
\begin{figure}[t]
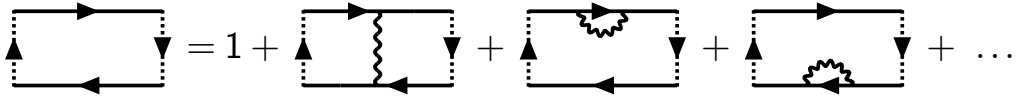

\Large
\begin{displaymath}
 \Scat \hspace{1.2cm} \mathsf{ = 1\hspace{0.1cm}+ }\hspace{-0.3cm} 
 \ScatA \hspace{1.2cm} \mathsf{ +}\hspace{-0.3cm} 
 \ScatB \hspace{1.2cm} \mathsf{ +}\hspace{-0.3cm} 
 \ScatC \hspace{1.2cm}\mathsf{ + }\hspace{0.2cm}\ldots
\end{displaymath}
\caption[a]{\small 
Symbolic representation of the Wilson loop to $\mathcal{O}\mathsf(g^2)$. Temporal Wilson lines, corresponding to simple identity matrices, are represented by dotted lines. The end points of the time-ordered gluon propagator, represented by a wiggly line, are integrated along the solid lines.}\label{Calculation}
\end{figure}

\noindent
In analogy with potential models at vanishing temperature a potential $V(t,\vec{r})$ is defined by assuming the time evolution of the mesonic correlator,
\be
iC^{21}(t,\vec{r})\equiv\int d^3\vec{x} \left<\bar{\psi}(t,\vec{x}+\frac{\vec{r}}{2})\gamma^{\mu}W\psi(t,\vec{x}-\frac{\vec{r}}{2})\bar{\psi}(0,\vec{0})\gamma_{\mu}\psi(0,\vec{0})\right>,
\ee
to be governed by the following Schr\"odinger equation, which has been shown to remain valid in a thermal setting at leading order \cite{Beraudo:2007ky}:
\be\label{Schroedinger}
i\partial_tC^{21}(t,\vec{r})=\left[2M-\frac{\triangle_{\vec{r}}}{M}+V(t,\vec{r})\right]C^{21}(t,\vec{r}).
\ee
$M$ denotes the mass of the constituent quarks while $W$ represents a straight Wilson line connecting the quark fields at time t. The notation follows the usual conventions in the Schwinger-Keldysh formalism. $V(t,\textbf{r})$ is deliberately introduced as a complex quantity with the imaginary part parametrising damping effects induced by the medium, thus generalising the concept of a potential to a thermal setting. To parametrise the evolution of the correlator at large times the potential $V(\vec{r})$ is subsequently defined as the infinite time limit of $V(t,\vec{r})$. Focusing on infinitely heavy quarks the amplitude $C^{21}(t,\vec{r})$ is represented by a Wilson loop of spatial extent $r$ and temporal extent $t$ upon introduction of another point splitting. Ambiguities due to time and path ordering can be avoided by replacing the Wightman propagator with the time-ordered propagator $C^{11}(t,\vec{r})$ at positive times. The static potential, which is referred to as the \textit{real-time static potential}, is thus obtained from the relations
\be\label{Wilson Loop}
V(\vec{r})\equiv\lim_{t\rightarrow\infty}\frac{i\partial_t C^{11}(t,\vec{r})}{C^{11}(t,\vec{r})}\hspace{0.3cm}\textrm{and}\hspace{0.3cm}iC^{11}(t,\vec{r})\equiv\frac{1}{N}Tr <\hat{\mathcal{T}}\hspace{-0.4cm}\Scat\hspace{1.0cm}>.
\ee\\[-0.2cm]
The Wilson loop is assumed to be oriented in the z-direction, with $\hat{\mathcal{T}}$ indicating the time-ordering.  For convenience a temporal gauge is chosen fixing the temporal Wilson lines to unity as indicated by dotted lines. Summing up the series of diagrams (Figure \ref{Calculation}), resulting from an expansion of the Wilson loop to first order in $g^2$, the following expression is obtained for the real-time static potential,\\[-0.55cm]
\be
V(\vec{r})=\lim_{t\rightarrow\infty}i\partial_t C^{11}(t,\vec{r})=\lim_{t\rightarrow\infty} 2g^2C_F\int \frac{d^4k}{(2\pi)^4}e^{i\omega t}\frac{\omega}{(k^3)^2}(1-\cos{k^3r}) i\tilde{G}_{33}^{11}(k)
\ee\\[-0.2cm]
where $\tilde{\bf G}^{11}(k)$ is the Fourier transform of the time-ordered HTL gluon propagator, which is related to the retarded propagator $\tilde{\bf G}^R(k)$ and Wightman propagator $\tilde{\bf G}^{21}(k)$ via\\[-0.15cm]
\be
\tilde{\bf G}^{11}=\tilde{\bf G}^R+\tilde{\bf G}^{12}=\tilde{\bf G}^R+e^{-\beta\omega}\tilde{\bf G}^{21}.
\ee

\newpage\noindent
Using the identity
\be
i\pi\delta(\omega)=\lim_{t\rightarrow \infty}\frac{e^{i\omega t}}{\omega},
\ee
the real-time static potential is obtained from the static limits of the afore mentioned propagators,
\be\label{RTSP1}
V(\vec{r})=-g^2C_F\int \frac{d^3\vec{k}}{(2\pi)^3}\frac{1-\cos{k^3r}}{(k^3)^2}\lim_{\omega\rightarrow 0}\omega^2(\tilde{G}^R_{33}(\omega,\vec{k})+\tilde{G}^{21}_{33}(\omega,\vec{k})),
\ee
with a factor $\omega^2$ cancelling the quadratic divergency of the temporal gauge propagator. 
The retarded propagator $\tilde{\bf G}^R$ in temporal gauge is explicitly given by the expression
\begin{equation}
\tilde{G}_{ij}^R(k)=-\frac{1}{k^2-\Pi_T(k)}(\delta_{ij}-\frac{k_ik_j}{\vec{k^2}})-\frac{1}{\vec{k}^2-\Pi_L(k)}\frac{k_ik_j}{\omega^2}.
\end{equation}
The longitudinal and transversal self-energies,
\be
\Pi_L(k)=m_D^2\left[\frac{\omega}{2 |\vec{k}|}log\frac{\omega+|\vec{k}|}{\omega-|\vec{k}|}-1\right]~~\textrm{and}~
\Pi_T(k)=\frac{m_D^2}{2}\frac{\omega^2}{\vec{k}^2}\left[1-\frac{\omega^2-\vec{k}^2}{2\omega |\vec{k}|}log\frac{\omega+|\vec{k}|}{\omega-|\vec{k}|}\right],
\ee
develop an imaginary part for $\omega < |{\bf k}|$. The Wightman propagator is readily obtained from the retarded propagator by using the KMS condition 
$
\tilde{\bf G}^{21}(k)=-2 i (n_B(\omega)+1)Im {\tilde{\bf G}^R}(k),
$
which relates both propagators in thermal equilibrium ($n_B(\omega)$ is the Bose distribution). An analytic expression for the real-time static potential is finally obtained by inserting the static limits of the propagators into (\ref{RTSP1}), with the real part corresponding to the usual Debye-screened potential: 
\ba\label{RTSP}
V(\vec{r})&=&g^2C_F\int \frac{d^3\vec{k}}{(2\pi)^3}(1-\cos{k^3r})\left\{\frac{1}{\vec{k}^2+m_D^2}-i\frac{\pi m_D^2}{\beta}\frac{1}{|\vec{k}|(\vec{k}^2+m_D^2)^2}\right\}\nonumber\\
&=&-\frac{g^2C_F}{4\pi}\left[m_D+\frac{e^{-m_D r}}{r}\right]-i\frac{g^2C_F}{2\pi\beta}\int_0^{\infty}\frac{dz z}{(z^2+1)^2}\left[1-\frac{\sin{(m_D r z)}}{m_D r z}\right].
\ea 
\noindent
\begin{figure}[t]
\Large
\centering
\includegraphics[width=6cm]{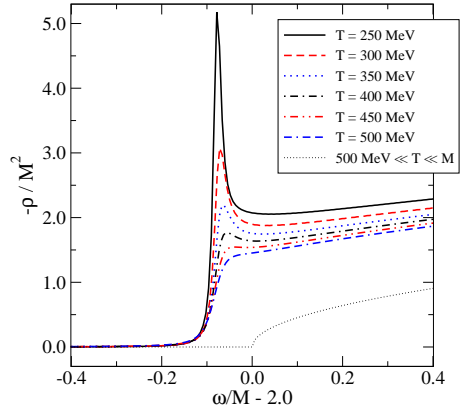}
\caption[a]{\small 
Melting of the bottomonium spectral function at finite temperature as obtained from the non-relativistic Schr\"odinger equation (\ref{Schroedinger}) using the real-time static potential. The figure is taken from \cite{Laine:2007gj}.}\label{SpectralFunction}
\end{figure}
The heavy-quarkonium spectral function $\rho(\omega)$ is obtained by reinserting the real-time static potential into the finite-mass Schr\"odinger equation (\ref{Schroedinger}) and employing the relation:
\be
\rho(\omega)\equiv\frac{1}{2}\left(1-e^{-\beta \omega}\right)\int_{-\infty}^{\infty}dt e^{i\omega t} iC^{21}(t,\vec{0}).
\ee
The spectral function for bottomonium \cite{Laine:2007gj} shows the expected widening of the resonance peak with increasing temperature which is induced by the imaginary part of the potential (Figure \ref{SpectralFunction}). Being the integral transform (\ref{RTSP1}) of the Wightman propagator, the imaginary part is associated with Landau damping. It is important to emphasise that, while the analytic continuation of the potential $V(i\beta,\vec{r})$ can be identified with the singlet free energy, the real-time static potential does not correspond to this quantity (see also \cite{Jahn:2004qr}). The static potential at finite temperature, as defined here from first principles, can in fact not be related in a straightforward way to an analogous quantity in euclidean space.
\section{Classical lattice gauge-theory simulations}
\begin{figure}[t]
\label{Comparison}
\Large
\centering
\includegraphics[width=5.5cm]{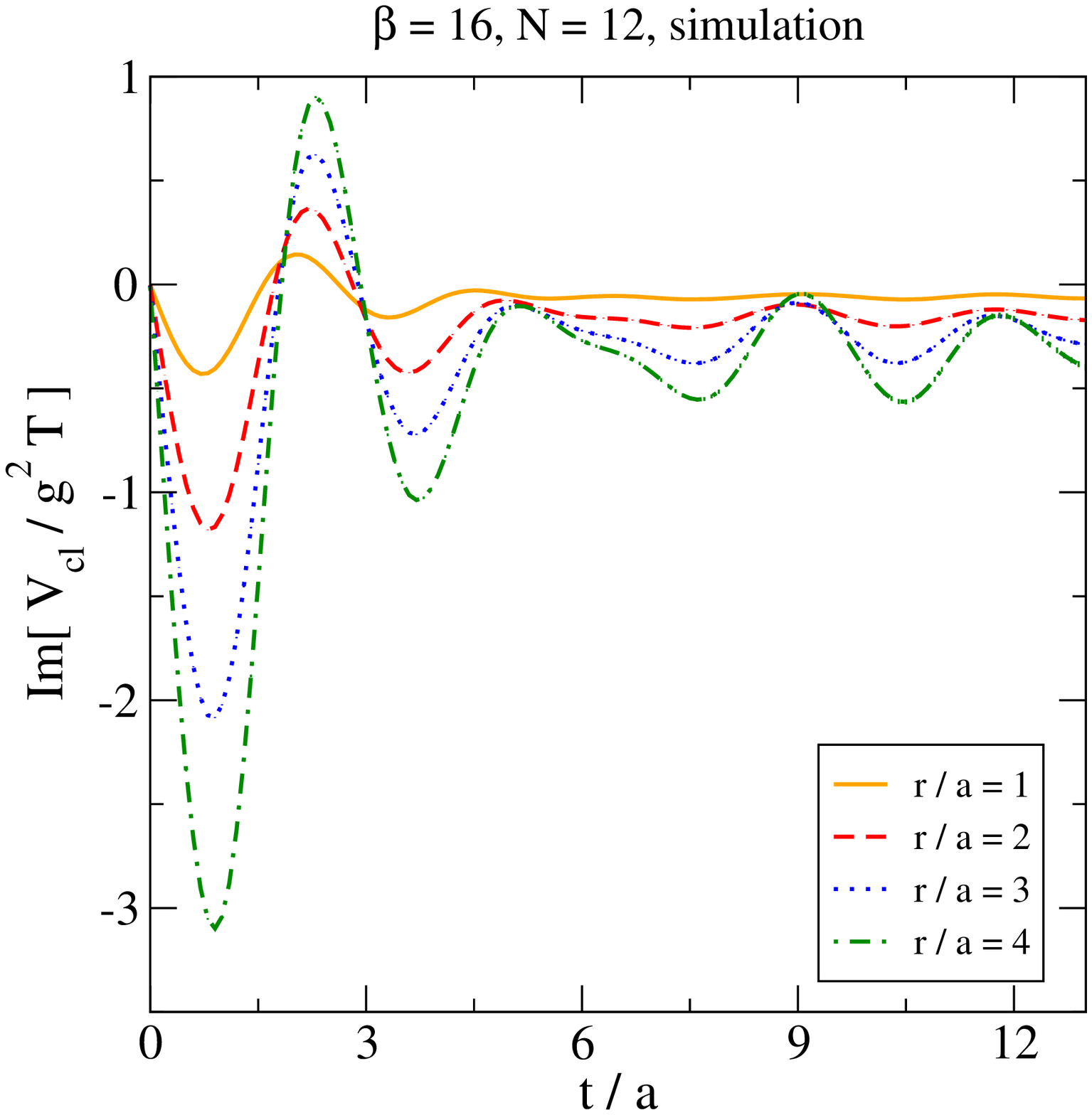}\includegraphics[width=5.5cm]{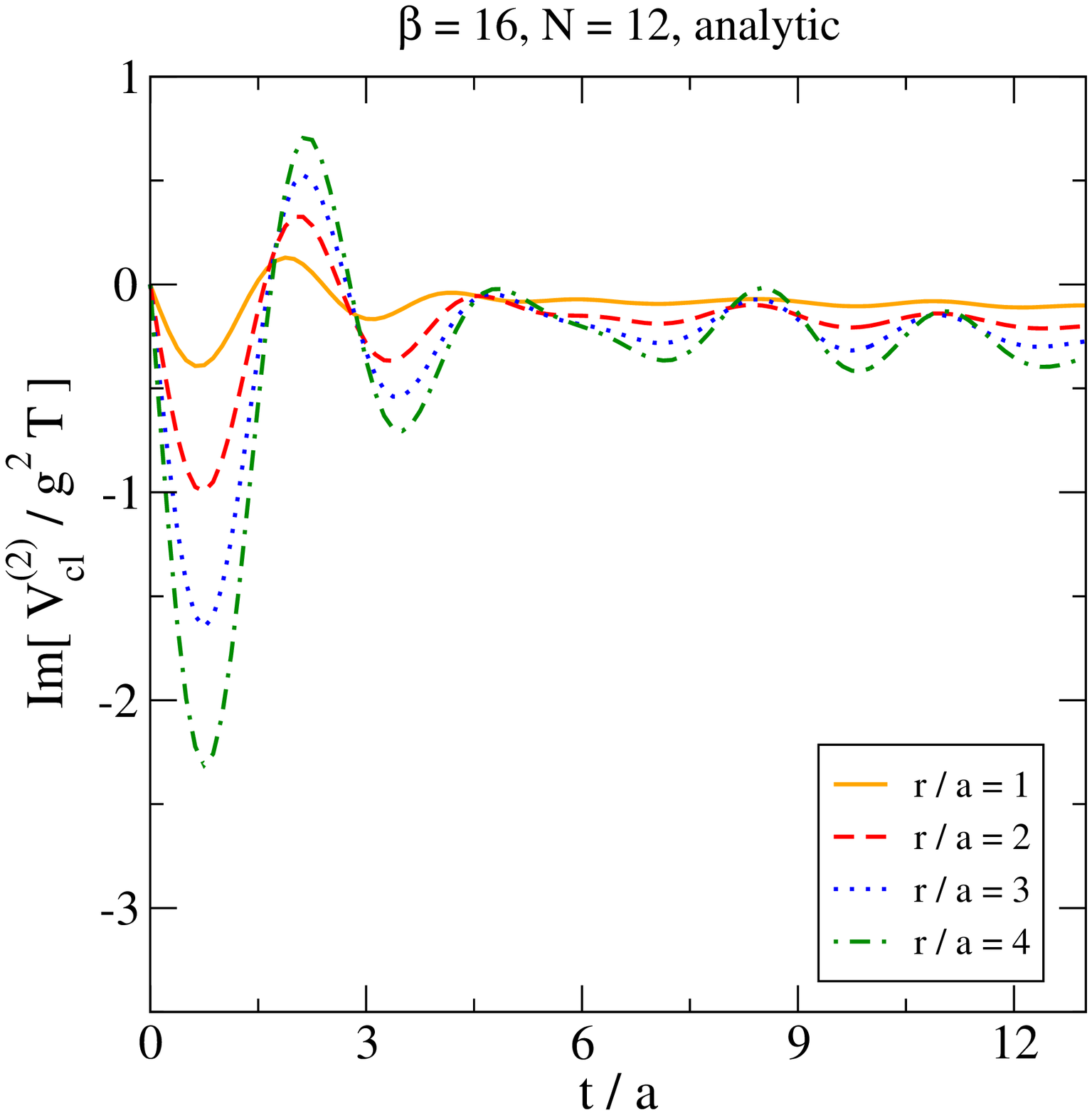}\\
\begin{small}
\begin{tabular}{|c|c|c|c|c|c|c|c|c|}
\hline
&$\beta$&N&$a m_D$&confs&$r=1a$&$r=2a$&$r=3a$&$r=4a$\\
\hline
Simulation&16.0&12&0.0&200&-0.060(2)&-0.156(8)&-0.246(26)&-0.319(56)\\
&16.0&16&0.0&160&-0.059(2)&-0.155(8)&-0.245(22)&-0.326(48)\\
&16.0&12&0.211&200&-0.059(2)&-0.147(7)&-0.229(23)&-0.297(51)\\
\hline
\hline
Analytic&16.0&$\infty$&0.0&-&-0.0601&-0.1145&-0.1507&-0.1737\\
\hline
\end{tabular}
\end{small}\\
\caption[a]{\small 
Time-dependence of the imaginary part of $V(t,{\bf r})$ as obtained from lattice-regularized perturbation theory and classical simulations. Below is an overview of the results in the large-time limit [3] for $\beta =16$. Results from classical and full HTL-improved simulations ($am_D > 0$) agree within error bars.}\label{LatticeResults}\vspace{-0.15cm}
\end{figure}

\noindent
Since the introduced static potential is a genuine real-time quantity, the extent to which physical results are subject to non-perturbative corrections needs to be assessed employing real-time lattice techniques. Formally the Yang-Mills-Vlasov equations are discretised on a 3-dimensional spatial lattice keeping the time coordinate continouus and choosing a temporal gauge. A review of the numerical implementation of these equations \cite{Laine:2007qy} is beyond the scope of this note and the discussion will therefore be restricted to the purely classical Yang-Mills theory. The partition function of the spatial lattice takes the following form,\\[-0.3cm]
\begin{equation}\label{partitionfunction}
 Z\equiv\int \! \mathcal{D}U_i\, \mathcal{D}E_i\, \delta(G) 
 e^{-\beta H}\;,\hspace{6mm}
 H\equiv\frac{1}{N_c}\sum_x\left[ \sum_{i<j} Tr (1-U_{ij})+\frac{1}{2} Tr( E_i^2 )\right]
 \;,
\end{equation}
with spatial links $U_i$ corresponding to discretised colour-magnetic fields and colour-electric fields defined via $\dot{U}_i=iE_iU_i$. The plaquette is denoted as $U_{ij}$. A discretised form of Gauss law is introduced via
$
G(x)\equiv \sum_i\left[E_i(x)-U_{-i}(x)E_i(x-\hat{i})U^+_{-i}(x)\right]=0
$. To evaluate the statistical expectation value of time dependent quantities, the partition function is supplemented by the evolution equations
\be
 \dot{U}_i(x)=iE_i(x)U_i(x) \;, \quad
 E_i = \sum_a E_i^a T^a \;, \quad
 \dot{E}_i^a(x)=- 2 Im Tr [T^a\sum_{|j|\neq i} U_{ij}(x)] 
 \;,
\ee
which follow from a variation of the discretised Yang-Mills action with respect to link variables. The thermalisation algorithm, generating the ensemble of configurations according to the statistical weight appearing in (\ref{partitionfunction}), is summarised as follows:
\begin{enumerate}
\item Pre-generate the spatial gauge links $U_i$ with a 3d Monte Carlo simulation.\vspace{-2mm}
\item Generate the electric fields from a gaussian distribution [cf.\ eq.~(\ref{partitionfunction})].\vspace{-2mm}
\item Project onto the space of physical configurations, satisfying the Gauss law.\vspace{-2mm}
\item Evolve the fields using the EOM, and repeat from step 2, until the fields have thermalised.
\end{enumerate}
For details on the implementation of the full Hard-Loop improved theory see \cite{Laine:2007qy}. Focusing on the imaginary part of the potential, which remains present in the classical limit,  a rectangular Wilson loop  of spatial extent $r=|\vec{r}|$ and temporal extent $t$ was measured using classical  and HTL-improved\newpage\noindent lattice simulations. In Figure \ref{Comparison} the time dependence of the correlator is compared to a lattice-regularised perturbative result \cite{Laine:2007qy}. Non-perturbative corrections were found to amplify the imaginary part of the potential, which is extracted according to (\ref{Wilson Loop}) in the large time limit, by up to $100 \%$, widening the quarkonium peak in Figure 2 but leaving the qualitative structure unchanged.

\vspace{-0.2cm}

\section{Conclusions}
\noindent
The non-relativistic Schr\"odinger equation used in potential models for quarkonia at zero temperature is generalised to a thermal setting. The corresponding \textit{real-time static potential} is derived in leading order perturbation theory from first principles, with an imaginary part inducing the melting of the quarkonium spectral function expected at finite temperature. Non-perturbative processes present in the classical limit of finite-temperature gauge theory amplify the imaginary part of the potential by up to $100\%$ and thus introduce an additional widening of the quarkonium resonance.

\end{document}